\documentclass[final,3p,times]{elsarticle}

\usepackage{graphicx}
\usepackage{amssymb}
\usepackage{color}
\usepackage{url}
\usepackage{multirow}
\usepackage{amsmath}
\usepackage{dsfont}

\journal{Physica A}

\begin{document}

\begin{frontmatter}

\title{The cooling-off effect of price limits in the Chinese stock markets}
\author[SS,RCE]{Yu-Lei Wan}
\author[hnu1,hnu2]{Gang-Jin Wang}
\author[RCE,BS]{Zhi-Qiang Jiang}
\author[RCE,BS]{Wen-Jie Xie}
\author[SS,RCE,BS]{Wei-Xing Zhou \corref{cor1}}
\cortext[cor1]{Corresponding author. Address: 130 Meilong Road, P.O. Box 114, School of Business,
              East China University of Science and Technology, Shanghai 200237, China, Phone: +86-13564271495.}
\ead{wxzhou@ecust.edu.cn} %
\address[SS]{Department of Mathematics, East China University of Science and Technology, Shanghai 200237, China}
\address[RCE]{Research Center for Econophysics, East China University of Science and Technology, Shanghai 200237, China}
\address[hnu1]{College of Business Administration, Hunan University, Changsha 410082, China}
\address[hnu2]{Center of Finance and Investment Management, Hunan University, Changsha 410082, China}
\address[BS]{Department of Finance, East China University of Science and Technology, Shanghai 200237, China}

\begin{abstract}
The price limit trading rule is one of the most widely adopted measures on restricting stock price volatilities in some stock markets. It is expected to stabilize the stock markets and enhance the efficiency of the market allocations. The existence of the cooling-off effect or the magnet effect, induced by the price limit trading rule, is the main controversy of this policy. In this paper, we investigate the cooling-off effect (opposite to the magnet effect) from two aspects. Firstly, from the viewpoint of dynamics, we study the existence of the cooling-off effect by following the dynamical evolution of some financial variables over a period of time before the stock price hits its limit. Secondly, from the probability perspective, we investigate, with the logit model, the existence of the cooling-off effect through analyzing the high-frequency data of all A-share common stocks traded on the Shanghai Stock Exchange and the Shenzhen Stock Exchange from 2000 to 2011 and inspecting the trading period from the opening phase prior to the moment that the stock price hits its limits. A comparison is made of the properties between up-limit hits and down-limit hits, and the possible difference will also be compared between bullish and bearish market state by dividing the whole period into three alternating bullish periods and three bearish periods. We find that the cooling-off effect emerges for both up-limit hits and down-limit hits, and the cooling-off effect of the down-limit hits is stronger than that of the up-limit hits. The difference of the cooling-off effect between bullish period and bearish period is quite modest. Moreover, we examine the sub-optimal orders effect, and infer that the professional individual investors and institutional investors play a positive role in the cooling-off effects. All these findings indicate that the price limit trading rule exerts a positive effect on maintaining the stability of the Chinese stock markets.
\end{abstract}
\end{frontmatter}

\section{Introduction}

The price limit trading rule, as a widely used method in some stock markets, aims to prevent the excessive fluctuations of stock prices. The limit price is usually set as a fixed fluctuated percentage of the previous trading day's closing price. The price limit trading rule is expected to reduce the volatility of the stock prices and have a cooling-off effect on the stock market \cite{Ma-Rao-Sears-1989-JFSR}. However, the rule may cause a magnet effect at the same time, which means that the limit price acts as a magnet to attract more trades to concentrate around the limit. The magnet effect will result in a higher trading intensity and stronger price volatility when the price is close to the limit price \cite{Subrahmanyam-1994-JF}. From the behavioral side, the magnet effect occurs when the traders are in fear of the lack of liquidity and the possible position lock caused by imminent price limit hits. In this case, the traders choose to protect themselves by submitting aggressive sub-optimal orders, which usually induces wide price variations and heavy trading volumes.

Previous literature offers controversial empirical evidence on the effectiveness of the price limit trading rule. A large number of papers have been published about the effectiveness of the price limit trading rule \cite{Wan-Xie-Gu-Jiang-Chen-Xiong-Zhang-Zhou-2015-PLoS1}. Here, we refer to the most recent papers about it with the aim of providing new evidence about the price limit trading rule. Li et al. examined the effectiveness, cause and impact of price limits by comparing cross-listed stocks on the Chinese stock markets, Hong Kong stock market and New York stock market. They found the price limits have some effectiveness in preventing price discovery; however, the price limit is ineffective in either volatility spillover or trading interference \cite{Li-Zheng-Chen-2014-JIFMIM}. Similarly, Lu examined the effectiveness of the price limit trading rule by using cross-listed stocks in the Chinese stock markets and Hong Kong stock market and found that the influence of price limits becomes weaker as limit-hitting stocks are traded more actively. They also concluded that excessive trading activities on individual stocks delay the process of price discovery and aggravate the volatility spillover \cite{Lu-2016-JSJTU}. Zhang et al. investigated the inter-day effects of price limits policies that are employed in agent-based simulations. The trading mechanisms in this market are the same as those in China's stock markets. The results of these simulations demonstrate that both upper and lower price limits can cause a volatility spillover effect and a trading interference effect \cite{Zhang-Ping-Zhu-Li-Xiong-2016-PLoS1}. Wu et al. provided a detailed analysis of the price dynamics after the hits of up-limit or down-limit \cite{Wu-Wang-Li-2017-PA,Wu-Wang-Li-2018-PA}. They showed that the expected return is found to be a ``W'' shape, which reveals high probability of a continuous price limit hit on the following day. Lin et al. investigated whether and how price limits are related to the cross-section of stock return by using the data from the Taiwan Exchange. They showed that the value premium is stronger among stocks with lower limit-hit frequency \cite{Lin-Ko-Lin-Yang-2017-PBFJ}.  Chen et al. focused on the futures price distribution to investigate how to set an appropriate daily margin level in Taiwan. They concluded that the legal margin for single stock futures set at 13.5\% by the Taiwan Futures Exchange to prevent default risk appears too broad \cite{Chen-Chou-Fung-Tse-2017-RQFA}. Shams et al. studied the existence of the magnet effect on the Tehran stock exchange and also investigated the role of the active investors in the stock market. They found that the role of institutional investors in the magnet effect is more significant than that of non-institutional investors \cite{Shams-Mohammadi-Kordlouie-Mahdavirad-2014-IJEF}.  However, contradictory conclusions are still in existence. Deb et al. provided new evidence on the effectiveness of the price limit trading rule with the data from the Tokyo Stock Exchange over a period of 5 years from January 2001 to December 2005. They concluded that the price limit trading rule works quite efficiently for lower limit hits as there is no evidence of volatility spill-over \cite{Deb-Kale-Marisetty-2016-PBFJ}.

\section{Descriptions of the Chinese stock markets and Data sources}

The Shanghai Stock Exchange and the Shenzhen Stock Exchange are the two stock exchanges in mainland China. There are 1374 A-shares companies listed on the Shanghai Stock Exchange, and 2766 A-shares companies listed on the Shenzhen Stock Exchange by 2017/12/05. The Shanghai Stock Exchange (SHSE) was established on 1990/11/26 and started its operation on 1990/12/19. Shortly afterwards, the Shenzhen Stock Exchange (SZSE) was established on 1990/12/01, and started its operation on 1991/07/03. The range of price limits changed several times on both Shanghai Stock Market and Shenzhen Stock Market. In the case of the Shanghai Stock Exchange, the price limits were $\pm1\%$ in the initial operation stage, shortly afterwards, the price limits were $\pm0.5\%$. Between 1992/5/21 and 1996/12/15, the Shanghai Stock Exchange canceled the price limit trading rule. Similar changes have occurred on the Shenzhen Stock Exchange. Since 1996/12/16, the price limits are $\pm10\%$ for all common stocks and $\pm5\%$ for specially treated (ST and ST*) stocks traded on both two exchanges.

Our data come from RESSET (http://resset.cn/), which includes all A-share stocks traded on the Shanghai Stock Exchange and the Shenzhen Stock Exchange. The length of time in our sample is a total of 12 years which covers the period from 2000/01/04 to 2011/12/30. The lengths of time on different stocks vary due to different IPO dates on stocks. The quote frequency is about 5s before 2011/6/27 and 3s afterwards.

The Shanghai Stock Exchange Composite (SSEC) Index is a representative measure of the status in Chinese stock markets.
In our last paper, we divide the time period from 2000 to 2011 into alternating periods of bullish and bearish states. The stock market was bearish during the three time periods: 2001/6/14 - 2005/6/03, 2007/10/17 - 2008/10/27, and 2009/08/05 - 2011/12/30. Other periods like 2000/1/4- 2001/6/13, 2005/06/04 - 2007/10/16, and 2008/10/28 - 2009/08/04 are recognized as being bullish\cite{Wan-Xie-Gu-Jiang-Chen-Xiong-Zhang-Zhou-2015-PLoS1}.

For each stock, the up-limit and the down-limit on every trading day is determined as follows. Let $P_s(T)$ denote the closing price of stock $s$ on day $T$. The up-limit $P^+_s(T+1)$ and the down-limit $P^-_s(T+1)$ of stock $s$ on day $T+1$ are determined as $ P^{\pm}_s(T+1) = {\mathcal{R}\left[100P_s(T)(1\pm r)\right]}/100$. $\mathcal{R}\left[x\right]$ is a round operator of $x$ such that the daily price limits are rounded to the tick size (the tick size of all stocks is 0.01 Chinese Yuan) according to the {\textit{Trading Rules of Shanghai Stock Exchange}} (2003, 2006) and the {\textit{Shenzhen Stock Exchange Trading Rules}} (2003, 2006). $r$ is 10\%, 5\% specifically for common stocks and special treatment stocks.

\section{The logit model and explanatory variables}

\subsection{The logit model}

We investigate the magnet effect from the angle of probability with the logit model. If there exists the magnet effect, the conditional probability of price increasing (decreasing) will increase significantly when the stock price approaches its limits. Here we define the logit model referred in \cite{Hsieh-Kim-Yang-2009-JEF} and \cite{Zhang-Zhu-2014-cnJCQUT} as follows,
\begin{equation}
\log(\frac{{P({Y_k} = 1|{X_k})}}{{1 - P({Y_k} = 1|{X_k})}}) = X_k^{'}B
\label{Eq:X:k:B}
\end{equation}
 $Y_{k}$ for up-limit models is defined as follows,
\begin{equation}
\label{Eq:logit:regression:upmodel:Yk:definition}
 Y_k = 1,~{\mathrm{if}}~\left\{ \begin{array}{l}
 p_k > p_{k-1}\\
 p_k = p_{k-1} ~~ {\mathrm{and}}~~ p_k > \frac{1}{2}(a_k +  b_k)
\end{array} \right.
\end{equation}
Else $Y_{k}=0$. where $p_{k}$ denotes the trading price of the ${(k)}$-th trade, $a_{k}$ and $b_{k}$ denote respectively the best ask and bid prices of the ${(k)}$-th trade. Creatively, here we also define $Y_k=1$ when $ p_k = p_{k-1}, p_k > \frac{1}{2}(a_k +  b_k)$. Since there may exist trades of large sizes in the best ask (bid) price level, and only two or more serially buyer-initiated (seller-initiated) trades can consume these trades, we define $Y_k=1$ under this circumstance.

Similarly, $Y_{k}$ for down-limit models is defined as follows,
\begin{equation}
 \label{Eq:logit:regression:downmodel:Yk:definition}
 Y_k = 1,~{\mathrm{if}}~\left\{ \begin{array}{l}
 p_k < p_{k-1}\\
 p_k = p_{k-1} ~~ {\mathrm{and}}~~ p_k < \frac{1}{2}(a_k +  b_k)
 \end{array} \right.
\end{equation}
Else $Y_{k}=0$.

In the up-limit (down-limit) model, $odds$ is defined as the ratio between the probability of the stock price rising up (falling down) and the probability of price falling down (rising up),
\begin{equation}
 \label{Eq:logit:regression:model:odds}
 odds = \frac {P(Y_{k}=1 | X_{k})} {1-P(Y_{k}=1 | X_{k})}
\end{equation}

For each limit-hitting day of stock $s$, we run the logit regressions with all trades from the opening till hitting its first limits to investigate the cooling-off effect (magnet effect). The logit regressions discussed here do not take into account the limit-hitting days which attach its limits at the opening, on account of no price movement before limits under this circumstance. Firstly, we define some explanatory variables investigated in the logit model.

\subsection{Definition of explanatory variables}

Explanatory variables of the logit model defined here are mainly referred in \cite{Hsieh-Kim-Yang-2009-JEF} and \cite{Zhang-Zhu-2014-cnJCQUT}, specific definitions of explanatory variables are shown as follows.
The first variable is the {\textit{pre-event trading direction}} $IBS_{k-l}$, which is the direction of the $l$-th trade lagging behind the $k$-th trade. For up-limit hits we have
\begin{equation}
 \label{Eq:IBS:k:direction:up}
 IBS_{k-l} = \left\{ \begin{array}{rll}
 1,&~ {\mathrm{if}}\;~{p_{k-l}} > (a_{k-l} +b_{k-l})/2\\
-1,&~ {\mathrm{if}}\;~{p_{k - l}} \leqslant  (a_{k-l} +b_{k-l})/2
 \end{array} \right.
\end{equation}
and for down-limit hits we have
\begin{equation}
\label{Eq:IBS:k:direction:down}
IBS_{k-l} = \left\{ \begin{array}{rll}
 1,&~ {\mathrm{if}}\; ~ {p_{k-l}} < (a_{k-l} +b_{k-l})/2\\
-1,&~ {\mathrm{if}}\;~{p_{k - l}} \geqslant  (a_{k-l} +b_{k-l})/2
\end{array} \right.,
\end{equation}
where $p_{k-l}$ denotes the price of the order that triggers the ${(k-l)}$-th trade, and $a_{k-l}$ and $b_{k-l}$ denote respectively the best ask and bid prices right before the ${(k-l)}$-th trade. In other words, $IBS_{k-l}=1$ if the trading price of the order is greater than the average of quoted best bid and ask prices.

The second variable is the {\textit{size}} ${V_{k}}$ of the $k$-th trade, which is the natural logarithm of the transaction size of the $k$-th trade.

The third variable investigated is the trade-by-trade {\textit{yield}} or return, which is defined as follows
\begin{equation}
 yield_{k}= \log{p_{k}}-\log{p_{k-1}}.
 \label{Eq:Prehit:yield}
\end{equation}
The trade-by-trade {\textit{volatility}} is thus
\begin{equation}
 volatility_{k}= |yield_{k}|= |\log{p_{k}}-\log{p_{k-1}}|,
 \label{Eq:Prehit:volatility}
\end{equation}
which is the fourth variable.

The fifth variable investigated is the {\textit{bid-ask spread}} right before the $k$-th trade, an indicator of liquidity, which is defined as follows
\begin{equation}
 spread_{k} = \frac{2(a_k - b_k)}{(a_{k} + b_{k})}.
 \label{Eq:Prehit:spread}
\end{equation}

The sixth financial variable is the limit-order book {\textit{depth}}, another liquidity indicator, which is defined as follows
\begin{equation}
 depth_{k} = {\rm{sign}}\left[\sum\limits_{j = 1}^J {I_{u,d}\left(b_k^{j} V_k^{b,j} - b_k^{j} V^{a,j}\right)}\right] \log\left|\sum\limits_{j = 1}^J {I_{u,d}\left(b_k^{j} V_k^{b,j} - a_k^{j} V^{a,j}\right)}\right|,
 \label{Eq:Prehit:depth}
\end{equation}
where $I_u=1$ for up-limit hits, $I_d=-1$ for down-limit hits, ${\rm{sign}}[x]$ is an indicator function of $x$ such that ${\rm{sign}}[x]=1$ when $x>0$ and ${\rm{sign}}[x]=-1$ when $x<0$, $a_k^{j}$ and $b_k^{j}$ denote respectively the $j$-level ask and bid prices right before the $k$-th trade arrives, $V_k^{a,j}$ and $V_k^{b,j}$ denote the corresponding outstanding volumes on the $j$-th ask and bid price levels right before the $k$-th trade arrives, and $J$ is the number of levels of the limit order book that is visible to traders (hence $J=3$ before 5 December 2003 and $J=5$ afterwards).

\subsection{Summary statistics of five financial quantities}

Table~\ref{Tb:Statistics:financialvariables:prehit:16tick} digests the dynamics of five financial variables (trade size in lots, yield, volatility, bid-ask spread and limit-order book depth in lots) along the last 16 trades before limit hits by showing the mean and median values. Panel A presents the results for up-limit hits and Panel B for down-limit hits. We also separate limit hits during bullish and bearish market states. Note that $k=16$ corresponds to the earliest trade before the limit hit and $k=1$ corresponds to the last trade that is closest to the limit hit.

\setlength\tabcolsep{1.5pt}
\begin{table}[!ht]
  \centering
  \small
 \medskip
 \centering
\caption{\label{Tb:Statistics:financialvariables:prehit:16tick}
Summary statistics of five important financial quantities along the last 16 trades right before limit hits.}
\begin{tabular}{ccccccccccccccccccccccccccccccccccccccccc}
\hline
\hline
\multicolumn{22}{l}{{\small{Panel A: Up-limit }}}\\
\hline
\multirow{2}{*}{$k$} && \multicolumn{4}{c}{$V_k$} && \multicolumn{4}{c}{$yield_k(\times 10^{-3})$}  && \multicolumn{4}{c}{$volatility_k(\times 10^{-3})$} && \multicolumn{4}{c}{$spread_k(\times 10^{-3})$}&& \multicolumn{4}{c}{$depth_k$ }\\
\cline{3-6}\cline{8-11}\cline{13-16}\cline{18-21}\cline{23-26}
&&\multicolumn{4}{c}{Bullish~~Bearish}&& \multicolumn{4}{c}{Bullish~~Bearish} &&\multicolumn{4}{c}{Bullish~~Bearish}&&\multicolumn{4}{c}{Bullish~~Bearish}&&\multicolumn{4}{c}{Bullish~~Bearish}\\
\hline
16&&\multicolumn{4}{c}{10.05(10.17)~~10.16(10.28)} &&\multicolumn{4}{c}{0.64(0.00)~~1.03(0.00)} &&\multicolumn{4}{c}{1.35(0.00)~~1.79(0.44)} &&\multicolumn{4}{c}{2.13(1.56)~~2.09(1.32)} &&\multicolumn{4}{c}{3.79(2.76)~~4.18(2.78)}\\
15&&\multicolumn{4}{c}{9.94(10.05)~~9.95(10.06)} &&\multicolumn{4}{c}{0.86(0.00)~~1.20(0.00)} &&\multicolumn{4}{c}{1.54(0.50)~~1.86(0.65)} &&\multicolumn{4}{c}{2.14(1.60)~~2.05(1.34)} &&\multicolumn{4}{c}{4.11(2.76)~~4.64(2.78)}\\
14&&\multicolumn{4}{c}{9.81(9.94)~~9.77(9.88)} &&\multicolumn{4}{c}{0.92(0.00)~~1.13(0.00)} &&\multicolumn{4}{c}{1.66(0.71)~~1.83(0.76)} &&\multicolumn{4}{c}{2.16(1.62)~~2.01(1.34)} &&\multicolumn{4}{c}{4.50(2.76)~~5.01(2.78)}\\
13&&\multicolumn{4}{c}{9.71(9.83)~~9.62(9.71)} &&\multicolumn{4}{c}{0.98(0.00)~~1.09(0.00)} &&\multicolumn{4}{c}{1.69(0.82)~~1.80(0.80)} &&\multicolumn{4}{c}{2.15(1.64)~~1.95(1.34)} &&\multicolumn{4}{c}{4.81(2.76)~~5.33(2.78)}\\
12&&\multicolumn{4}{c}{9.60(9.72)~~9.47(9.59)} &&\multicolumn{4}{c}{0.96(0.00)~~1.00(0.00)} &&\multicolumn{4}{c}{1.67(0.85)~~1.69(0.79)} &&\multicolumn{4}{c}{2.17(1.66)~~1.95(1.35)} &&\multicolumn{4}{c}{5.11(2.76)~~5.58(2.77)}\\
11&&\multicolumn{4}{c}{9.50(9.60)~~9.38(9.47)} &&\multicolumn{4}{c}{0.93(0.00)~~0.96(0.00)} &&\multicolumn{4}{c}{1.68(0.89)~~1.68(0.82)} &&\multicolumn{4}{c}{2.16(1.66)~~1.92(1.35)} &&\multicolumn{4}{c}{5.40(2.76)~~5.76(2.77)}\\
10&&\multicolumn{4}{c}{9.44(9.54)~~9.28(9.39)} &&\multicolumn{4}{c}{0.93(0.00)~~0.90(0.00)} &&\multicolumn{4}{c}{1.68(0.90)~~1.64(0.84)} &&\multicolumn{4}{c}{2.17(1.69)~~1.90(1.35)} &&\multicolumn{4}{c}{5.63(2.76)~~5.94(2.77)}\\
9&&\multicolumn{4}{c}{9.38(9.48)~~9.20(9.31)} &&\multicolumn{4}{c}{0.89(0.00)~~0.87(0.00)} &&\multicolumn{4}{c}{1.65(0.89)~~1.60(0.82)} &&\multicolumn{4}{c}{2.17(1.69)~~1.89(1.34)} &&\multicolumn{4}{c}{5.83(2.76)~~6.08(2.77)}\\
8&&\multicolumn{4}{c}{9.31(9.43)~~9.15(9.25)} &&\multicolumn{4}{c}{0.86(0.00)~~0.81(0.00)} &&\multicolumn{4}{c}{1.63(0.90)~~1.54(0.82)} &&\multicolumn{4}{c}{2.16(1.68)~~1.86(1.34)} &&\multicolumn{4}{c}{6.02(2.76)~~6.22(2.77)}\\
7&&\multicolumn{4}{c}{9.28(9.39)~~9.08(9.19)} &&\multicolumn{4}{c}{0.85(0.00)~~0.76(0.00)} &&\multicolumn{4}{c}{1.62(0.89)~~1.49(0.81)} &&\multicolumn{4}{c}{2.16(1.70)~~1.83(1.33)} &&\multicolumn{4}{c}{6.17(2.76)~~6.34(2.77)}\\
6&&\multicolumn{4}{c}{9.23(9.34)~~8.99(9.09)} &&\multicolumn{4}{c}{0.83(0.00)~~0.72(0.00)} &&\multicolumn{4}{c}{1.60(0.88)~~1.47(0.79)} &&\multicolumn{4}{c}{2.15(1.70)~~1.85(1.34)} &&\multicolumn{4}{c}{6.31(2.76)~~6.43(2.77)}\\
5&&\multicolumn{4}{c}{9.19(9.30)~~8.95(9.05)} &&\multicolumn{4}{c}{0.78(0.00)~~0.71(0.00)} &&\multicolumn{4}{c}{1.59(0.88)~~1.44(0.77)} &&\multicolumn{4}{c}{2.15(1.69)~~1.80(1.32)} &&\multicolumn{4}{c}{6.48(2.77)~~6.49(2.76)}\\
4&&\multicolumn{4}{c}{9.16(9.26)~~8.89(8.97)} &&\multicolumn{4}{c}{0.76(0.00)~~0.64(0.00)} &&\multicolumn{4}{c}{1.55(0.86)~~1.40(0.75)} &&\multicolumn{4}{c}{2.14(1.69)~~1.80(1.32)} &&\multicolumn{4}{c}{6.58(2.77)~~6.54(2.76)}\\
3&&\multicolumn{4}{c}{9.12(9.23)~~8.87(8.94)} &&\multicolumn{4}{c}{0.75(0.00)~~0.64(0.00)} &&\multicolumn{4}{c}{1.55(0.87)~~1.40(0.77)} &&\multicolumn{4}{c}{2.12(1.68)~~1.80(1.32)} &&\multicolumn{4}{c}{6.66(2.76)~~6.61(2.76)}\\
2&&\multicolumn{4}{c}{9.08(9.20)~~8.83(8.93)} &&\multicolumn{4}{c}{0.70(0.00)~~0.61(0.00)} &&\multicolumn{4}{c}{1.53(0.85)~~1.36(0.73)} &&\multicolumn{4}{c}{2.11(1.68)~~1.78(1.31)} &&\multicolumn{4}{c}{6.74(2.77)~~6.68(2.76)}\\
1&&\multicolumn{4}{c}{9.10(9.21)~~8.77(8.84)} &&\multicolumn{4}{c}{0.69(0.00)~~0.59(0.00)} &&\multicolumn{4}{c}{1.50(0.84)~~1.34(0.71)} &&\multicolumn{4}{c}{2.10(1.68)~~1.76(1.31)} &&\multicolumn{4}{c}{6.75(2.76)~~6.70(2.75)}\\
\hline
\multicolumn{22}{l}{{\small{Panel B: Down-limit }}}\\
\hline
\multirow{2}{*}{$k$} && \multicolumn{4}{c}{$V_k$} && \multicolumn{4}{c}{$yield_k(\times 10^{-3})$}  && \multicolumn{4}{c}{$volatility_k(\times 10^{-3})$} && \multicolumn{4}{c}{$spread_k(\times 10^{-3})$}&& \multicolumn{4}{c}{$depth_k$ }\\
\cline{3-6}\cline{8-11}\cline{13-16}\cline{18-21}\cline{23-26}
&&\multicolumn{4}{c}{Bullish~~Bearish}&& \multicolumn{4}{c}{Bullish~~Bearish} &&\multicolumn{4}{c}{Bullish~~Bearish}&&\multicolumn{4}{c}{Bullish~~Bearish}&&\multicolumn{4}{c}{Bullish~~Bearish}\\
\hline
16&&\multicolumn{4}{c}{9.15(9.22)~~8.43(8.45)} &&\multicolumn{4}{c}{-0.59(0.00)~~-0.51(0.00)} &&\multicolumn{4}{c}{1.26(0.00)~~1.27(0.00)} &&\multicolumn{4}{c}{2.30(1.57)~~2.38(1.53)} &&\multicolumn{4}{c}{-9.82(-14.21)~~-10.39(-14.12)}\\
15&&\multicolumn{4}{c}{9.10(9.21)~~8.44(8.49)} &&\multicolumn{4}{c}{-0.70(0.00)~~-0.62(0.00)} &&\multicolumn{4}{c}{1.41(0.00)~~1.37(0.00)} &&\multicolumn{4}{c}{2.34(1.64)~~2.38(1.56)} &&\multicolumn{4}{c}{-9.08(-14.07)~~-9.74(-13.99)}\\
14&&\multicolumn{4}{c}{9.03(9.16)~~8.37(8.41)} &&\multicolumn{4}{c}{-0.84(0.00)~~-0.72(0.00)} &&\multicolumn{4}{c}{1.50(0.38)~~1.45(0.26)} &&\multicolumn{4}{c}{2.40(1.69)~~2.37(1.58)} &&\multicolumn{4}{c}{-8.20(-13.90)~~-8.96(-13.84)}\\
13&&\multicolumn{4}{c}{8.99(9.07)~~8.31(8.34)} &&\multicolumn{4}{c}{-0.84(0.00)~~-0.71(0.00)} &&\multicolumn{4}{c}{1.57(0.68)~~1.49(0.49)} &&\multicolumn{4}{c}{2.44(1.72)~~2.38(1.60)} &&\multicolumn{4}{c}{-7.28(-13.66)~~-8.27(-13.65)}\\
12&&\multicolumn{4}{c}{8.92(9.03)~~8.23(8.26)} &&\multicolumn{4}{c}{-0.85(0.00)~~-0.72(0.00)} &&\multicolumn{4}{c}{1.61(0.79)~~1.48(0.51)} &&\multicolumn{4}{c}{2.45(1.75)~~2.42(1.63)} &&\multicolumn{4}{c}{-6.45(-13.46)~~-7.54(-13.47)}\\
11&&\multicolumn{4}{c}{8.87(8.98)~~8.18(8.18)} &&\multicolumn{4}{c}{-0.84(0.00)~~-0.70(0.00)} &&\multicolumn{4}{c}{1.62(0.81)~~1.47(0.55)} &&\multicolumn{4}{c}{2.46(1.75)~~2.41(1.64)} &&\multicolumn{4}{c}{-5.64(-13.21)~~-6.93(-13.27)}\\
10&&\multicolumn{4}{c}{8.79(8.86)~~8.10(8.13)} &&\multicolumn{4}{c}{-0.85(0.00)~~-0.65(0.00)} &&\multicolumn{4}{c}{1.61(0.80)~~1.48(0.58)} &&\multicolumn{4}{c}{2.46(1.79)~~2.41(1.65)} &&\multicolumn{4}{c}{-4.97(-12.97)~~-6.27(-13.08)}\\
9&&\multicolumn{4}{c}{8.79(8.85)~~8.05(8.03)} &&\multicolumn{4}{c}{-0.82(0.00)~~-0.65(0.00)} &&\multicolumn{4}{c}{1.62(0.84)~~1.50(0.59)} &&\multicolumn{4}{c}{2.46(1.79)~~2.43(1.67)} &&\multicolumn{4}{c}{-4.36(-12.72)~~-5.74(-12.87)}\\
8&&\multicolumn{4}{c}{8.74(8.82)~~8.01(8.00)} &&\multicolumn{4}{c}{-0.78(0.00)~~-0.65(0.00)} &&\multicolumn{4}{c}{1.60(0.83)~~1.48(0.64)} &&\multicolumn{4}{c}{2.45(1.79)~~2.44(1.67)} &&\multicolumn{4}{c}{-3.77(-12.50)~~-5.11(-12.65)}\\
7&&\multicolumn{4}{c}{8.71(8.77)~~7.98(8.00)} &&\multicolumn{4}{c}{-0.80(0.00)~~-0.62(0.00)} &&\multicolumn{4}{c}{1.57(0.82)~~1.47(0.61)} &&\multicolumn{4}{c}{2.44(1.79)~~2.45(1.68)} &&\multicolumn{4}{c}{-3.33(-12.23)~~-4.60(-12.41)}\\
6&&\multicolumn{4}{c}{8.67(8.73)~~7.95(8.00)} &&\multicolumn{4}{c}{-0.73(0.00)~~-0.61(0.00)} &&\multicolumn{4}{c}{1.56(0.81)~~1.43(0.59)} &&\multicolumn{4}{c}{2.44(1.79)~~2.45(1.69)} &&\multicolumn{4}{c}{-2.86(-12.01)~~-4.16(-12.19)}\\
5&&\multicolumn{4}{c}{8.66(8.76)~~7.92(7.90)} &&\multicolumn{4}{c}{-0.74(0.00)~~-0.59(0.00)} &&\multicolumn{4}{c}{1.60(0.84)~~1.44(0.63)} &&\multicolumn{4}{c}{2.44(1.80)~~2.44(1.68)} &&\multicolumn{4}{c}{-2.39(-11.76)~~-3.74(-11.99)}\\
4&&\multicolumn{4}{c}{8.63(8.69)~~7.90(7.93)} &&\multicolumn{4}{c}{-0.72(0.00)~~-0.54(0.00)} &&\multicolumn{4}{c}{1.54(0.83)~~1.42(0.60)} &&\multicolumn{4}{c}{2.41(1.80)~~2.43(1.69)} &&\multicolumn{4}{c}{-2.01(-11.48)~~-3.41(-11.81)}\\
3&&\multicolumn{4}{c}{8.61(8.68)~~7.87(7.90)} &&\multicolumn{4}{c}{-0.69(0.00)~~-0.59(0.00)} &&\multicolumn{4}{c}{1.54(0.81)~~1.42(0.57)} &&\multicolumn{4}{c}{2.40(1.80)~~2.42(1.69)} &&\multicolumn{4}{c}{-1.71(-11.23)~~-3.17(-11.67)}\\
2&&\multicolumn{4}{c}{8.58(8.64)~~7.85(7.86)} &&\multicolumn{4}{c}{-0.67(0.00)~~-0.52(0.00)} &&\multicolumn{4}{c}{1.52(0.83)~~1.39(0.57)} &&\multicolumn{4}{c}{2.40(1.79)~~2.43(1.69)} &&\multicolumn{4}{c}{-1.47(-11.03)~~-2.96(-11.54)}\\
1&&\multicolumn{4}{c}{8.55(8.61)~~7.82(7.82)} &&\multicolumn{4}{c}{-0.63(0.00)~~-0.50(0.00)} &&\multicolumn{4}{c}{1.51(0.82)~~1.41(0.60)} &&\multicolumn{4}{c}{2.41(1.79)~~2.45(1.70)} &&\multicolumn{4}{c}{-1.19(-10.76)~~-2.77(-11.44)}\\
  \hline\hline
\end{tabular}
\end{table}

We find that the mean trade size $V_k$ decreases with decreasing $k$ before up-limit hits and down-limit hits during bullish and bearish market states, which suggests that the average trade size becomes smaller when the stock price approaches the limit. It might suggest that more retail traders are rushing into the market and reflect a herding behavior. The mean trade size is slightly larger in the bullish period than in the bearish period before up-limit hits (Panel A, except for $k=16$ and $k=15$) and down-limit hits (Panel B). During the same market state, the trades before up-limit hits have larger sizes than those before down-limit hits. These findings indicate that traders are more aggressive when the price is rising in the short run (approaching up-limits) and in the long run (bullish periods). The results are similar for the median of trade sizes.

The medians of trade-by-trade returns are 0 for all $k$'s, suggesting that more than 50\% transactions do not affect the price. The magnitude of mean return increases from $k=16$ to around $k=14$ and then decreases with decreasing $k$. With a few exceptions ($k=11$ to $k=16$ in Panel A), the magnitude with the same $k$ during bullish periods is larger than that during bearish periods. For the same market state, the magnitude is larger before up-limit hits than that before down-limit hits. These observations are reasonable due to the relationship between trade size and immediate price impact \citep{Lillo-Farmer-Mantegna-2003-Nature,Lim-Coggins-2005-QF,Zhou-2012-NJP,Zhou-2012-QF}.

The mean and median of volatility increases first and then decrease with decreasing $k$. The mean and median of volatility during bullish periods are both greater than their counterparts during bearish periods, when the trades are close to up-limit hits (Panel A) or down-limit hits (Panel B). During the same market state, the volatility is higher before up-limit hits than that before down-limit hits. The behavior of bid-ask spreads is quite similar to the volatility.

When $k$ decreases, the mean depth increases before up-limit hits during bullish and bearish periods. However, the median depth does not change during bullish periods and exhibits a slight downward tendency during bearish periods when the price approaches the up-limits. The mean and median values of the depth are slightly higher during bearish periods for large $k$'s and slightly lower during bearish periods for small $k$'s. The dynamics of the depth completely differs from that before down-limit hits. We find that the depth $|depth_k|$ decreases when the price approaches the down-limits. In addition, the depth $|depth_k|$ is larger during bearish periods than that during bullish periods.

\section{Evidence of the cooling-off effect}

\subsection{Main evidence}

We select explanatory variables mainly referred in \cite{Hsieh-Kim-Yang-2009-JEF} and \cite{Zhang-Zhu-2014-cnJCQUT}, the right term $X_k^{'}B$ in the logit model(see Eq.~(1)) are defined as follows,
\begin{equation}
 \label{Eq:logit:regression:model}
 \begin{split}
 X_k^{'}B = {\beta _0} + {\beta _1}\Delta {T_k}+ {\beta _2}V_{k-1}*IBS_{k-1} + {\beta _3}V_{k-2}*IBS_{k-2}+ {\beta _4}V_{k-3}*IBS_{k-3}\\
 + {\beta _5}{yield_{k-1}}+ {\beta _6}MKT_{k-1} + {\beta _7}MK{T_{k-2}} + {\beta _8}MK{T_{k-3}} + {\beta _9}spread_{k-1} + {\beta_{10}}depth_{k-1}
 \end{split}
\end{equation}
where $\Delta {T_k}$ denotes the duration (in seconds) between the $(k-1)$-th trade and the $k$-th trade which captures the duration effect. We denote $V_{k-1}*IBS_{k-1}$ as the directional trading volume which captures the per-unit volume impact of the buyer-initiated (seller-initiated) effect. Here we do not select $IBS_{k-1}$ which is used as the independent explanatory variable referred in \cite{Hsieh-Kim-Yang-2009-JEF} and \cite{Zhang-Zhu-2014-cnJCQUT}, because the logit regressions generate multi-collinearity in our logit regressions. $MKT_{k-l}$ denotes $l$ lags of one-minute logarithmic returns of the SSEC Index or the SZSE Component Index. For limit-hitting days, since we consider $MKT_{k-l}$ as one of our explanatory variables, the first trade of $k$ considered in our logit model is defined as the first trade occurring after 9:33 a.m. Besides, $yield_{k-1}$ captures trade-by-trade return effect on $odds$, and $spread_{k-1}$, and $depth_{k-1}$ captures liquidity effect.

We run 85725 logit regressions for all the limit-hitting days with each trade from the opening till hitting its first limits, Table~\ref{Tb:Statistics:logitmodel:estimates} shows the results of the logit regressions, including the numbers of significantly positive, negative, and insignificantly estimates at the 5\% level.

\setlength\tabcolsep{3pt}
\begin{table}[!ht]
  \centering
  \small
 \caption{\label{Tb:Statistics:logitmodel:estimates} Summary statistics of parameter estimations for the logit regressions. }
 \medskip
 \centering
\begin{tabular}{cccccccccccccccccccccccccccccccccccccccccccccc}
\hline\hline
   Period & Model  && $\beta_{0}$ & $\beta_{1}$ & $\beta_{2}$ & $\beta_{3}$ & $\beta_{4}$ & {$\beta_{5}$ } & $\beta_{6}$ & $\beta_{7}$ & $\beta_{8}$ & $\beta_{9}$ & $\beta_{10}$ & \\\hline
  Bullish & Up-limit  & $+$ &        1253& 1023& 7271& 1966& 8160& 179& 7742& 1792& 1163& 2180& 12403&\\
     &   (N=25070)   & $-$ &        4822& 757& 940& 7723& 346& 18290& 140& 247& 1030& 979& 134&\\
     &      & 0 &        18995&23290&16859&15381&16564&6601&17188&23031&22877&21911&12533&\\
     &      & Sign &        $-$& $+$ & $+$ & $-$ & $+$ & $-$ & $+$ & $+$ & $+$ & $+$ & $+$ &\\\hline
  Bearish & Up-limit  & $+$ &        1015& 962& 3505& 1539& 7519& 80& 7778& 2171& 1007& 2386& 11140&\\
     &  (N=22635)    & $-$ &        4383& 653& 1745& 5364& 224& 16326& 142& 266& 1540& 779& 135&\\
     &      & 0  &        17237&21020&17385&15732&14892&6229&14715&20198&20088&19470&11360&\\
     &      & Sign  &        $-$& $+$ & $+$ & $-$ & $+$ & $-$ & $+$ & $+$ & $-$ & $+$ & $+$ &\\\hline
  Bullish & Down-limit  & $+$ &        929& 1343& 3876& 646& 3352& 10999& 30& 136& 928& 1004& 106&\\
     &   (N=14753)   & $-$ &        2042& 123& 716& 5171& 195& 41& 8146& 1486& 769& 728& 5572&\\
     &      &0 &        11782&13287&10161&8936&11206&3713&6577&13131&13056&13021&9075&\\
     &      & Sign &        $-$& $+$ & $+$ & $-$ & $+$ & $+$ & $-$ & $-$ & $+$ & $+$ & $-$ &\\\hline
  Bearish & Down-limit  & $+$ &        995& 2233& 4059& 856& 5746& 17571& 38& 179& 1115& 3208& 258&\\
     &   (N=23267)   & $-$ &        5318& 245& 1402& 5257& 166& 14& 12184& 3154& 1672& 506& 7148&\\
     &  &0 &    16954&20789&17806&17154&17355&5682&11045&19934&20480&19553&15861&\\
     &      & Sign  &        $-$& $+$ & $+$ & $-$ & $+$ & $+$ & $-$ & $-$ & $+$ & $+$ & $-$ &\\
  \hline\hline
\end{tabular}
\end{table}

If there exists the magnet effect, the conditional probability of a price increase (decrease) would increase significantly when the stock price approaches its limits. In other words, we may observe significantly positive (negative) $\beta _5$ for up-limit (down-limit) models. However, we infer from Table~\ref{Tb:Statistics:logitmodel:estimates} that about 73.0\%, 72.1\% of $\beta _5$ are significantly negative for up-limit in bullish periods and bearish periods and 74.6\%,75.5\% of $\beta _5$ are significantly positive for down-limit in bullish periods and bearish periods. We calculate mean (median) values of $\beta _5$, specifically, -256.1 (-237.0), -272.5 (-248.0) in the two cases (up-limit in bullish periods, up-limit in bearish periods), and 240.7 (214.4), 215.1 (187.0) in the two cases (down-limit in bullish periods, down-limit in bearish periods). The logit model shows that under the condition of controlling other explanatory variables, we only consider the effect of $yield$ on $odds$. We observe that $odds$ respectively decreases $(e^{\beta _5/1000}-1)$, that is, 22.6\% (21.1\%), 23.9\% (22.0\%) for every increasing 0.1\% change of $yield$ in the two cases (up-limit in bullish periods, up-limit in bearish periods), and $(e^{-\beta _5/1000}-1)$, that is, 21.4\% (19.3\%), 19.4\% (17.1\%) for every decreasing 0.1\% change of $yield$ in the two cases (down-limit in bullish periods, down-limit in bearish periods). We conclude that there exists the cooling-off effect when the stock price approaches its limits from the angle of probability. Nevertheless, other explanatory variables included in Eq.~\ref{Eq:logit:regression:model} are mainly used to control possible determinants, and these explanatory variables are still worthy of discussion. Specifically, for the interaction term $V_{k-l}*IBS_{k-l}$, we find that $V_{k-1}*IBS_{k-1}$, $V_{k-3}*IBS_{k-3}$ have a more significantly positive effect on $odds$ than that of $V_{k-2}*IBS_{k-2}$. About $MKT_{k-l}$  which is defined as $l$ lags of one-minute logarithmic returns of the SSEC Index(SZSE Component Index), however, only $MKT_{k-1}$  has the significantly positive effect on $odds$ indicating the short-term memory of marketing effect. And $depth$ produces a significant effect only second to $yield$ on $odds$.

\subsection{Fitness test}

For each limit-hitting day $i$, we define $\hat{Y_k}$ as the estimated value of $Y_k$ predicted by the logit regressions. We further define $Y_k^{*}$ by virtue of the Classification Table as follows ,
\begin{equation}
 \label{Eq:logit:regression:model:Classification:Table}
 Y_{k}^{*} =  \left\{ \begin{array}{l}
 1,~~~ {\mathrm{if}}\; ~\hat{Y_k}\geq 0.5\\
 0,~~~ {\mathrm{else}}\
 \end{array} \right.\
\end{equation}

We define $Q_{i}$ as the total number of the trades from the opening till hitting its first limits on the limit-hitting day $i$, and $Q_i^{*}$ denotes the number of $Y_{k}$ correctly predicted by $Y_{k}^{*}$. We define the accuracy $A_{i}$ as the ratio of $Y_{k}$ been correctly predicted by $Y_{k}^{*}$,
\begin{equation}
 \label{Eq:logit:regression:model:accuracy}
 A_{i} = \frac{Q_i^{*}}{Q_i}
\end{equation}

\cite{Paul-Mcfaddan-1974} apply $\rho$-square measured with maximum likelihood estimation which is similar to $R$-square measured with OLS estimation in the linear regressions to checkout fitting effects of the logit regressions, and the $\rho$-square is defined as follows,

\begin{equation}
 \label{Eq:logit:regression:model:mcfaddan:r:square}
 \left\{ \begin{array}{l}
 \rho^{2}=1- \frac {L^{*}(\hat{\beta})} {L^{*}(\overline{\beta})} \\
 L^{*}(\beta)=\sum\limits_{q = 1}^{Q} \sum\limits_{k = 1}^{J_{q}} {f_{k,q}* \log {P_{k,q}(\beta)} }
 \end{array} \right.\
\end{equation}
where $L^{*}$ is the likelihood function, $\beta$ is the parameter vector at which the calculated probabilities $P_{k,q}(\beta)$ are being computed, $f_{k,q}$ is relative frequencies with repetitions, $\hat{\beta}$ is defined to be the maximum of the likelihood estimator , and $\overline{\beta}$ is zero or zero except for coefficients of alternative dummies.

\begin{figure}[htb]
 \centering
\includegraphics[width=6cm]{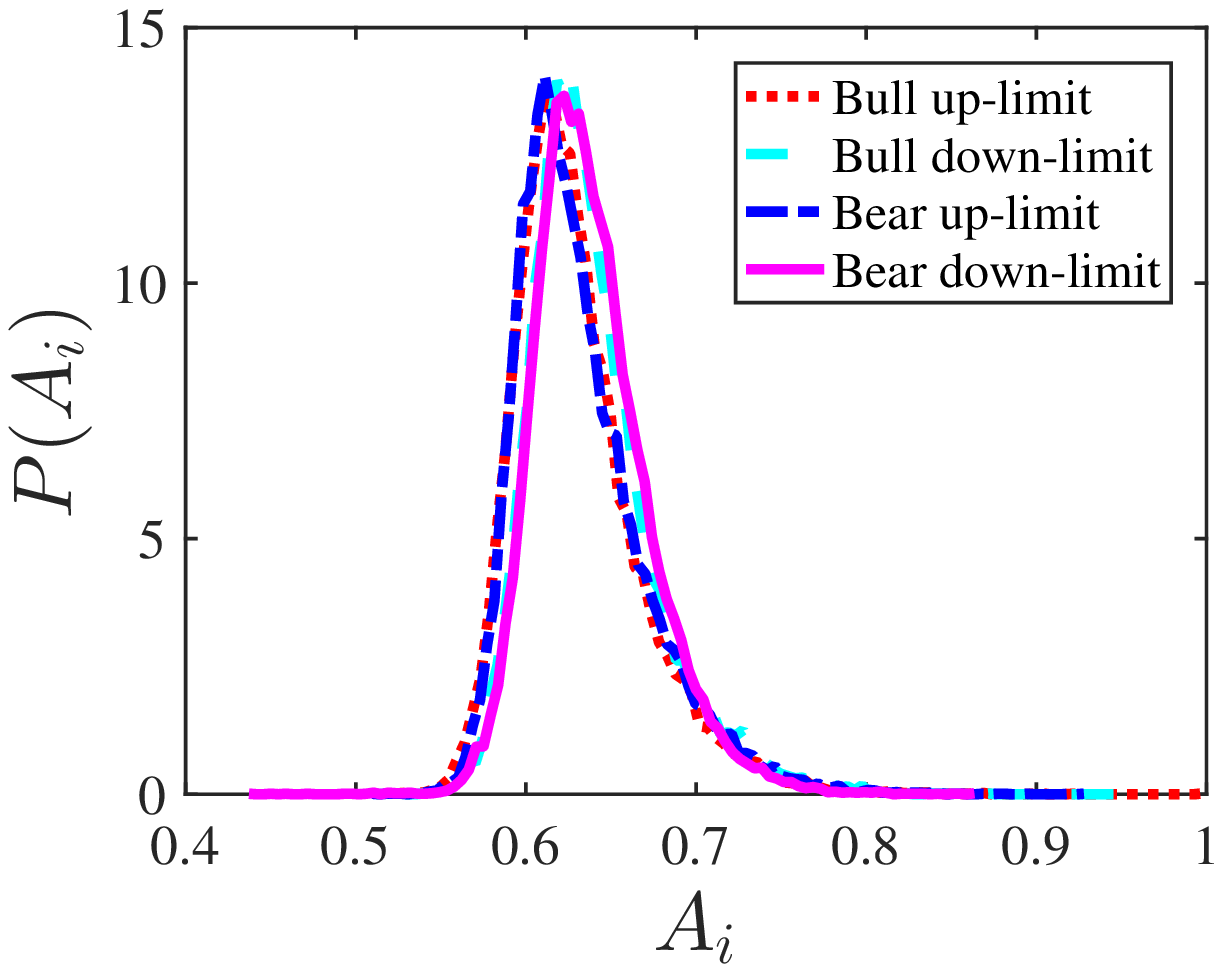}
   \hskip 0.5cm
 \includegraphics[width=6cm]{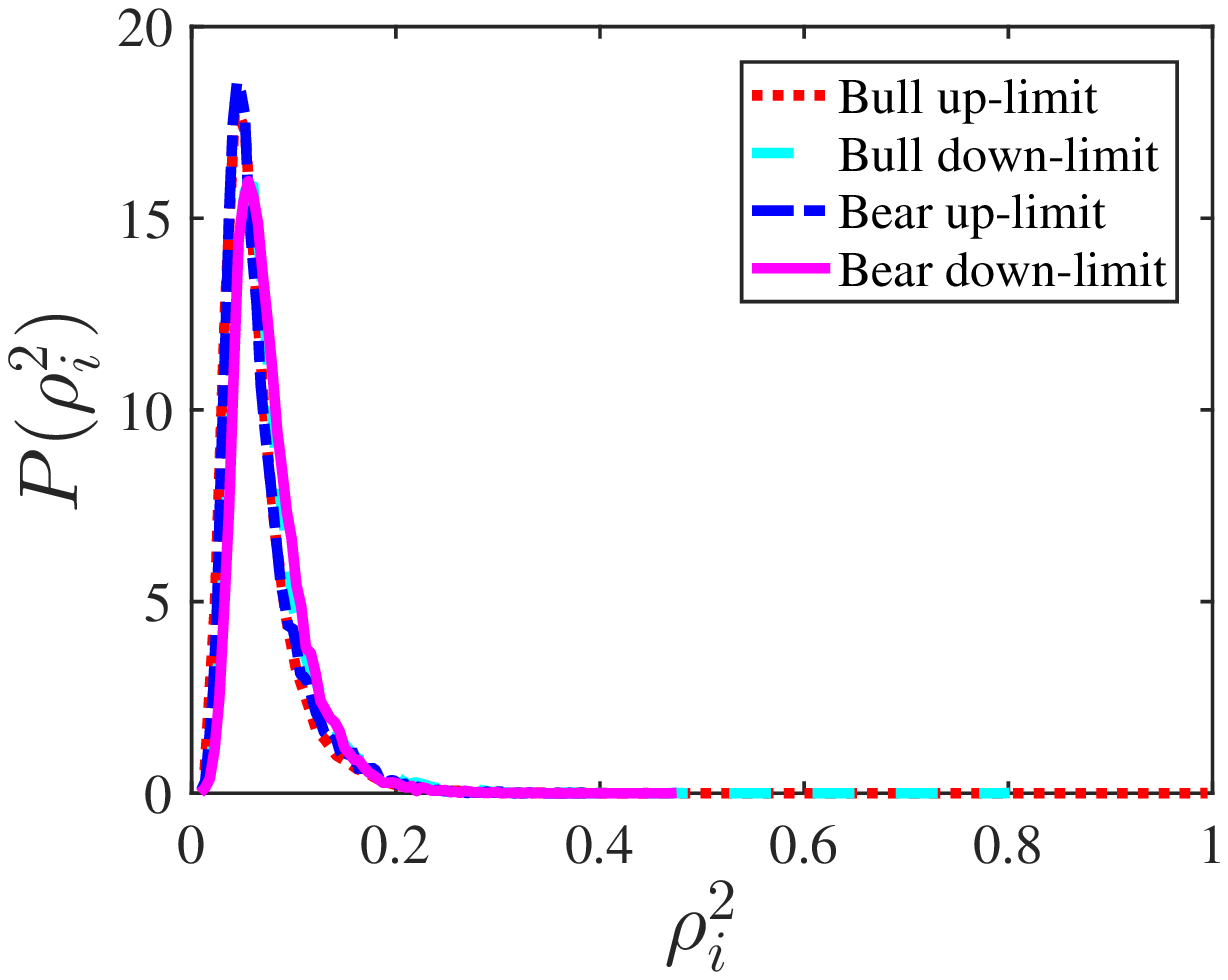}
 \caption{\label{Fig:logit:model:fitness} Probability density functions of the accuracy $A_i$ and $\rho_i^{2}$ in the four cases.}
\end{figure}

We calculate $\rho_i^2$ for each logit regression of the limit-hitting day $i$. Fig.~\ref{Fig:logit:model:fitness} shows the empirical distributions of $A_i$ and $\rho_i^{2}$ in the four cases (up-limit in bullish periods, down-limit in bullish periods, up-limit in bearish periods, and down-limit in bearish periods).

Although the $R$-square index is a more familiar concept in the OLS estimation, it is not as well behaved as the $\rho$-square measure with MLE estimation used in the logit regressions documented in \cite {Mcfaddan-1978-BTM} and \cite{Mcfaddan-Domencich-1975-UTD}. These papers also point out that the $\rho$-square index tends to be considerably lower than the $R$-square index, for example, the range interval [0.2, 0.4] of $\rho$-square used in the logit model represents excellent fit (equivalent to the range interval [0.7, 0.9] of $R$-square used in the linear regressions). We calculate the median (mean) values of $\rho^{2}$, specifically, 0.06 (0.08), 0.07 (0.07), 0.06 (0.07), 0.07 (0.08) in the four cases (up-limit in bullish periods, down-limit in bullish periods, up-limit in bearish periods, and down-limit in bearish periods). We also get the median (mean) values of $A$, specifically, 0.62 (0.63), 0.63 (0.64), 0.62 (0.63), 0.63 (0.63) in the four cases. We conclude that the logit model used here is reliable.

\subsection{Robustness test}

The probit regression is used as an alternative model for the robustness check. Similar to the logit model, the accuracy $A_{i}^{\rm{probit}}$ is defined as the ratio of $Y_{k}$ correctly predicted by $Y_{k}^{**}$ with the probit regression. We introduce the variable $\Delta A_{i}$, defined as the difference between $A_{i}^{\rm{logit}}$ and $A_{i}^{\rm{probit}}$.

\begin{figure}[htb]
 \centering
 \includegraphics[width=9cm]{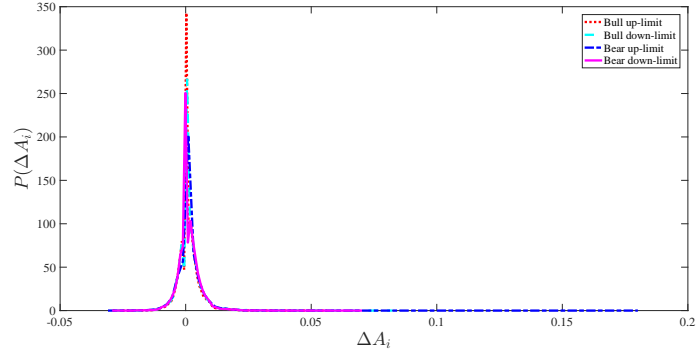}
 \caption{\label{Fig:logit:model:fitness:delta_accuracy}  Probability density functions of  $\Delta A_{i}$ in the four cases.}
\end{figure}

Fig.~\ref{Fig:logit:model:fitness:delta_accuracy} shows that the range interval of $\Delta A_{i}$ for all four cases is mainly concentrated in $[-0.02,0.04]$ and the four mean values of $\Delta A_{i}$ is close to zero, indicating that the results are approximately consistent between the logit model and the probit model. Moreover, the difference of $\rho_{i}^2$ between each logit regression and its corresponding probit regression processed on the limit-hitting day $i$ yields the identical result. These findings further support the evidence that the logit model is reliable for the sample data.

\subsection{Sub-optimal orders effect}

Subrahmanyam shows that `circuit breakers' (trading halts), by causing agents to suboptimally advance trades in time, may have the perverse effect of increasing price variability and exacerbating price movements \cite{Subrahmanyam-1994-JF}. In this section, we investigate the sub-optimal orders effect by first defining the $k$-th trade of the limit-hitting day $i$ to be the sub-optimal trade as follows,
\begin{equation}
 \label{Eq:sub-optimal:orders:definition}
IS_{k}  =\begin{cases}1 & {\mathrm{if }}~ p_{k} < b_k ~ or ~ p_{k}> a_k,\\0 & {\mathrm{else}}.\end{cases}.
\end{equation}
We also define the variable $W_{i}$ as the ratio of sub-optimal trades to total pre-hit trades on the limit-hitting day $i$, i.e.,
\begin{equation}
 \label{Eq:sub-optimal:orders:weight}
 W_i = \frac{{\sum\limits{IS_{k}}}} {S_i}
\end{equation}
where  $S_i$ denotes the numbers of trades from the opening till the price hitting its limits on the limit-hitting day $i$.

\begin{figure}[htb]
 \centering
 \includegraphics[width=7cm]{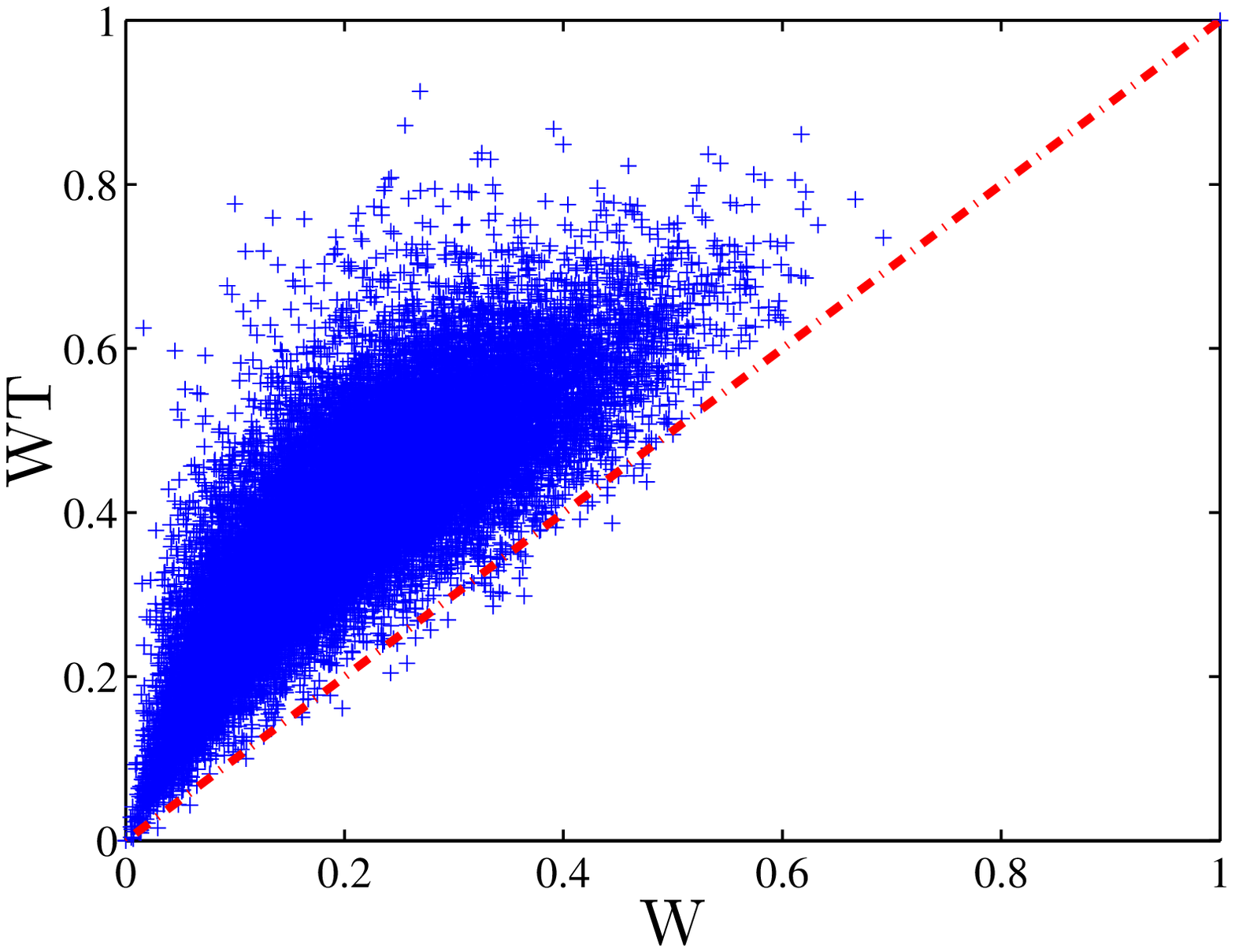}
 \includegraphics[width=7cm]{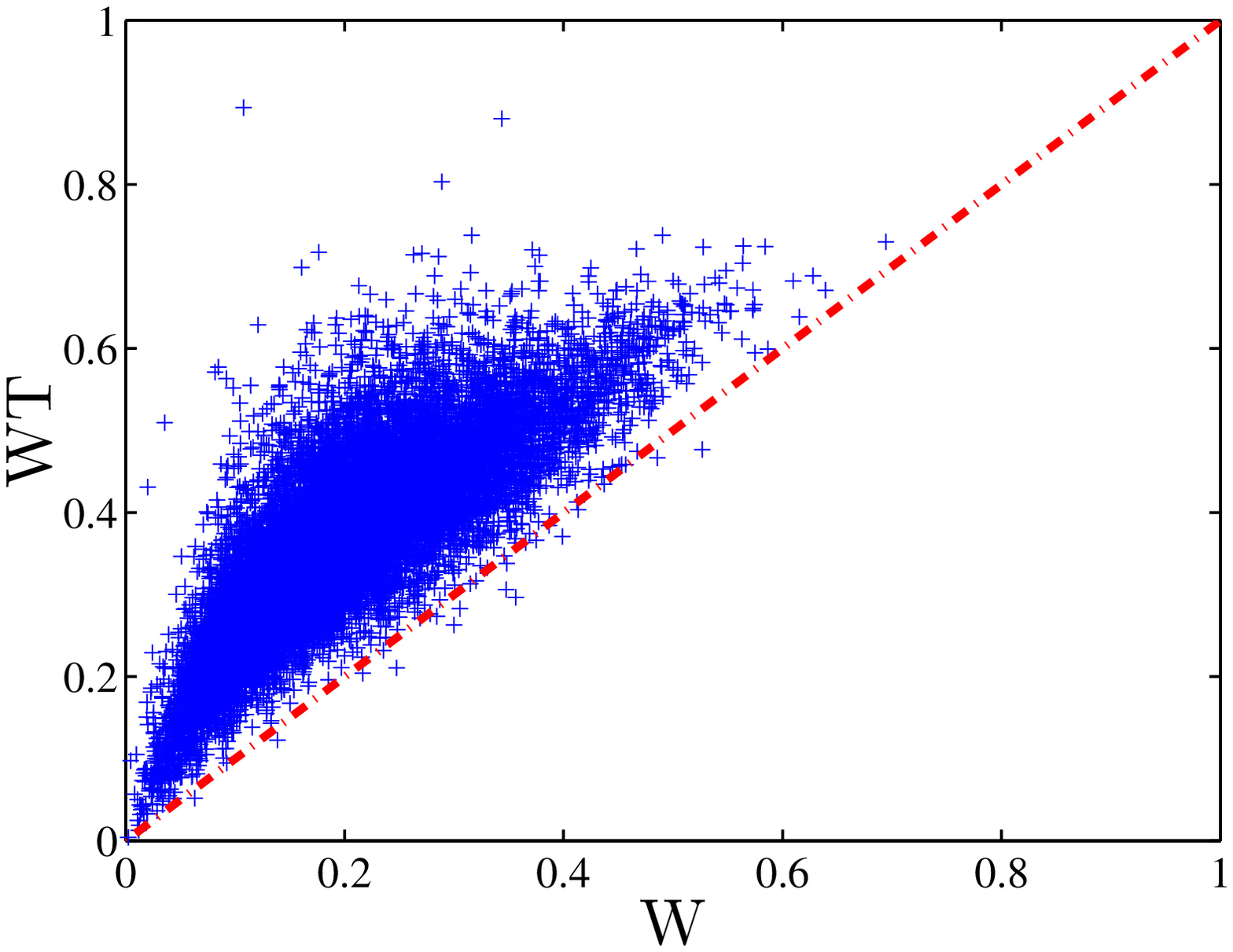}
 \includegraphics[width=7cm]{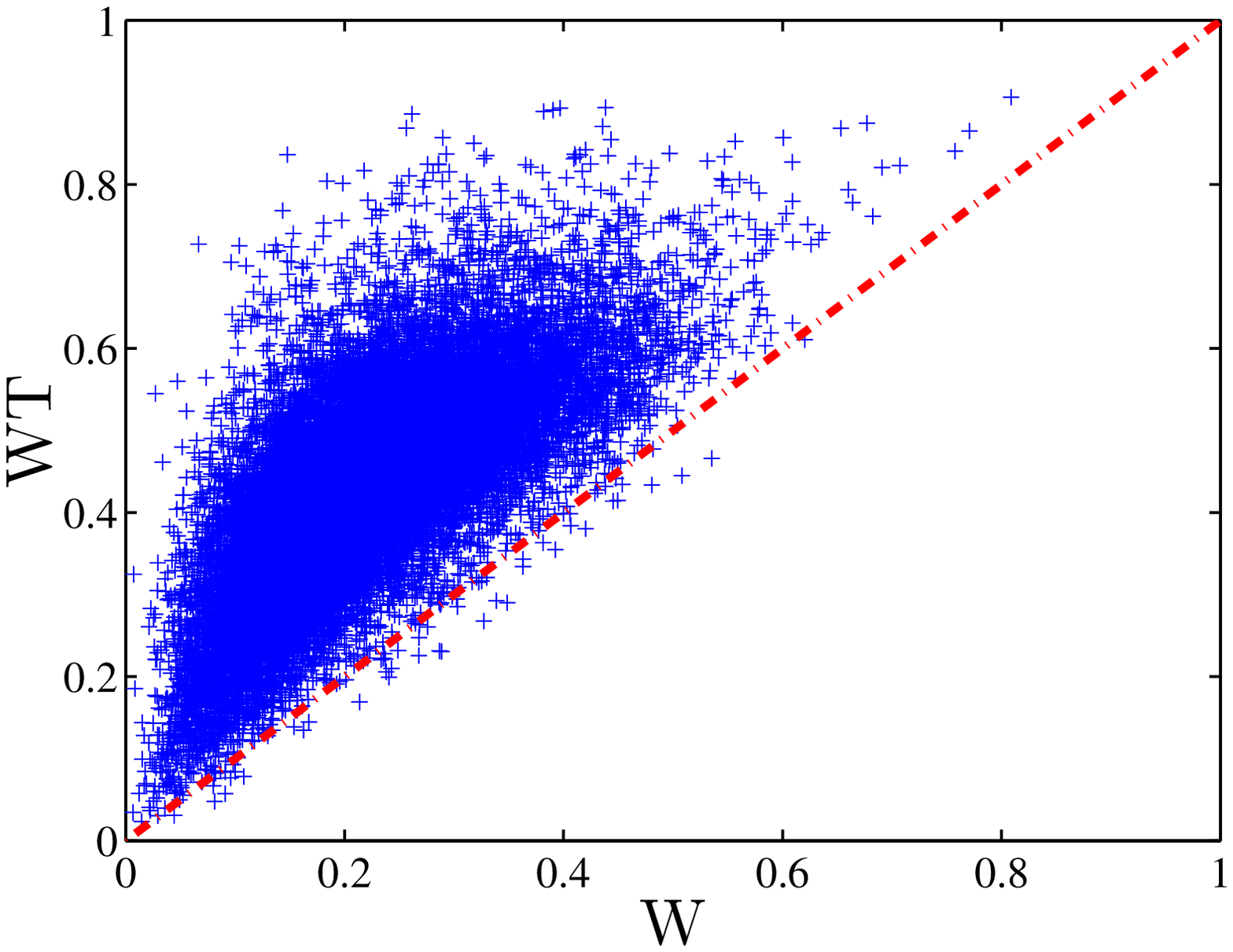}
 \includegraphics[width=7cm]{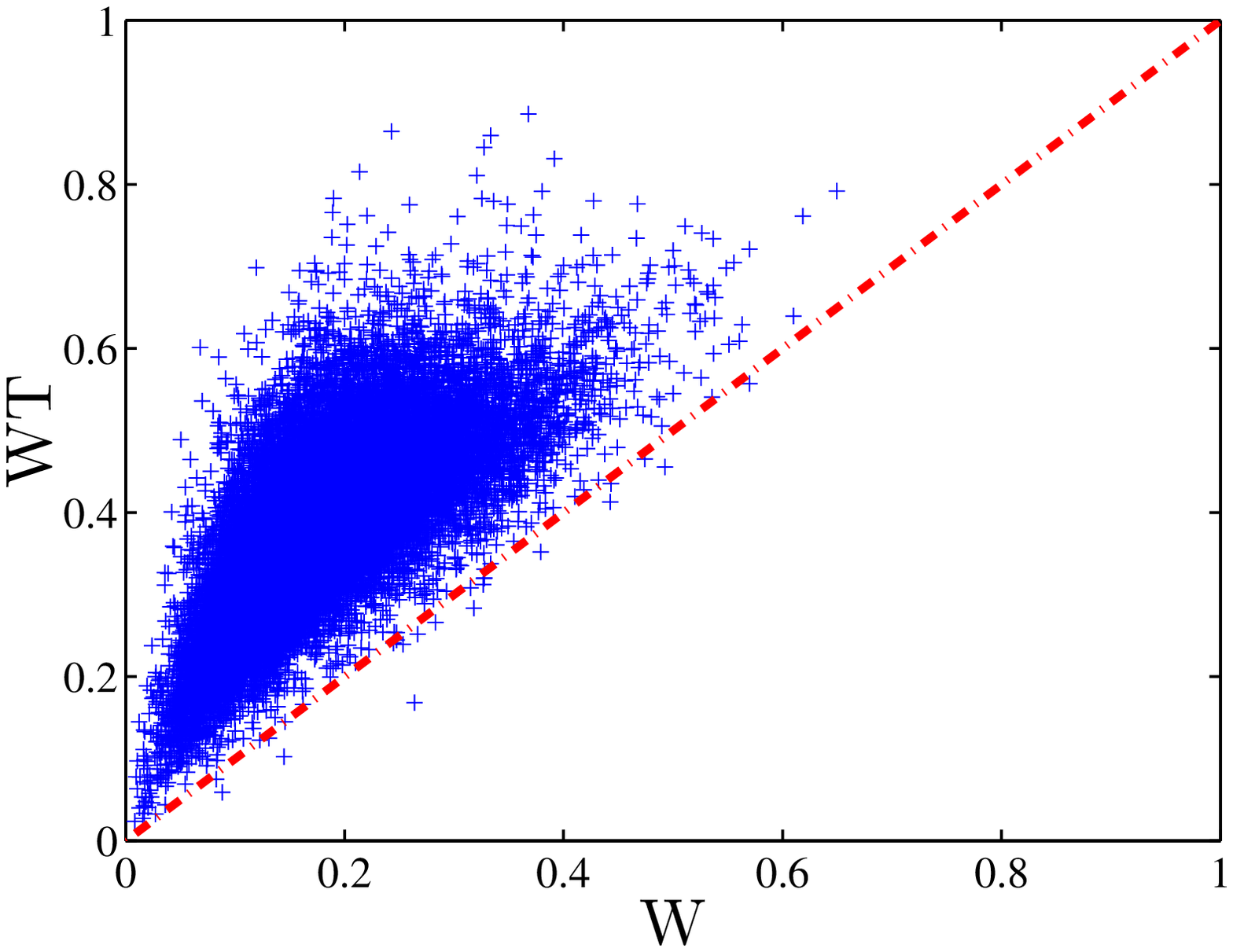}
 \caption{\label{Fig:Suboptimal:tick:weight} Top left (right) panel shows the relationship between $WT_{i}^{u}$ ($WT_{i}^{d}$) and $W_{i}^{u}$ ($W_{i}^{d}$) in the Bullish period. Bottom left (right) panel shows the relationship between $WT_{i}^{u}$ ($WT_{i}^{d}$) and $W_{i}^{u}$ ($W_{i}^{d}$) in the Bearish period. }
\end{figure}

We further introduce another variable $WT_{i}$, calculated as the ratio between the trading volume of sub-optimal trades and the trading volume of total pre-hit trades on the limit-hitting day $i$, i.e.,
\begin{equation}
 \label{Eq:sub-optimal:orders:trading:volume:weight}
 WT_i = \frac{{\sum\limits {V_{k}*IS_{k}}}} {\sum\limits {V_{k}}}
\end{equation}
where $V_k$ denotes the trading volume of the $k$-th trade on limit-hit day $i$. In Fig.~\ref{Fig:Suboptimal:tick:weight} we classify the relationships between  $WT_{i}$  and $W_{i}$ into four categories.
Fig.~\ref{Fig:Suboptimal:tick:weight} shows that although the sub-optimal trades only make up 20\%--40\% of the total trades, they contribute more to the total trading volumes.

We examine the logit model after taking into account the sub-optimal trades effect, and then $X_k^{'}B$ of the logit model is revised as follows,
\begin{equation}
 \label{Eq:logit:regression:model:suboptimal:build}
 \begin{split}
 X_k^{'}B = {\beta _0} + {\beta _1}\Delta {T_k}+ {\beta _2}{vol_{k-1}}*IBS_{k-1} + {\beta _3}{vol_{k-2}}*IBS_{k-2}+ {\beta _4}{vol_{k-3}}*IBS_{k-3}+ {\beta _5}{yield_{k-1}}\\
 + {\beta _6}{IS_{k-1}*yield_{k-1}}+{\beta _7}MKT_{k-1} + {\beta _8}MK{T_{k-2}} + {\beta _9}MK{T_{k-3}}+
 {\beta _{10}}spread_{k-1} + {\beta_{11}}depth_{k-1}
 \end{split}
\end{equation}

\setlength\tabcolsep{3pt}
\begin{table}[!ht]
  \centering
  \small
 \caption{\label{Tb:Statistics:logitmodel:estimates:suboptimal:ticks} Parameter estimations of the logit regressions considering the sub-optimal orders effect. }
 \medskip
 \centering
\begin{tabular}{cccccccccccccccccccccccccccccccccccccccccccccc}
\hline\hline
   Period & Model  && $\beta_{0}$ & $\beta_{1}$ & $\beta_{2}$ & $\beta_{3}$ & $\beta_{4}$ & $\beta_{5}$ & $\beta_{6}$ & $\beta_{7}$ & $\beta_{8}$ & $\beta_{9}$ & $\beta_{10}$ & $\beta_{11}$ & \\\hline
  Bullish & Up-limit  & $+$ &        1258&1014&7627&1839&8215&38&7758&7765&1824&1158&2197&12348&\\
     &   (N=25070)   & $-$ &        4775&767&835&8116&332&19093&205&139&257&1034&987&135&\\
     &      & 0 &        19037&23289&16608&15115&16523&5939&17107&17166&22989&22878&21886&12587&\\
     &      & Sign  &   $-$& $-$ & $+$ & $-$ & $+$ & $-$ & $+$ & $+$ & $-$ & $-$ & $+$ & $+$ &\\\hline
  Bearish & Up-limit  & $+$ &        1016&935&3734&1428&7573&11&5351&7794&2213&998&2413&11108&\\
     &   (N=22635)   & $-$ &        4375&668&1598&5740&221&16942&211&147&276&1549&775&145&\\
     &      & 0 &        17244&21032&17303&15467&14841&5682&17073&14694&20146&20088&19447&11382&\\
     &      & Sign  &     $-$& $+$ & $+$ & $-$ & $+$ & $-$ & $+$ & $+$ & $+$ & $-$ & $+$ & $+$ &\\\hline
  Bullish & Down-limit  & $+$ &        902&1340&4038&618&3395&11451&160&31&142&952&1022&5531&\\
     &   (N=14753)   & $-$ &        2056&126&646&5368&200&8&4491&8154&1515&775&725&102&\\
     &      & 0 &        11795&13287&10069&8767&11158&3294&10102&6568&13096&13026&13006&9120&\\
     &      & Sign  &    $+$& $+$ & $+$ & $-$ & $+$ & $+$ & $-$& $-$ & $-$ & $+$ & $+$ & $+$ &\\\hline
  Bearish & Down-limit  & $+$ &        991&2218&4228&822&5772&17938&325&37&181&1120&3220&7129&\\
     &   (N=23267)   & $-$ &        5338&249&1335&5448&168&5&4604&12168&3160&1692&518&261&\\
     &      & 0 &        16938&20800&17704&16997&17327&5324&18338&11062&19926&20455&19529&15877&\\
     &      & Sign  &     $+$& $+$ & $+$ & $-$ & $+$ & $+$ & $-$ & $-$ & $-$ & $-$ & $+$ & $+$ &\\
  \hline\hline
\end{tabular}
\end{table}

Table~\ref{Tb:Statistics:logitmodel:estimates:suboptimal:ticks} shows the results of the logit regressions considering the sub-optimal trades effect. We calculate the number of significantly positive, negative at 5\% level. $IS_{k-1}*yield_{k-1}$ captures the sub-optimal trades effect, significantly positive $\beta_6$ for up-hits and negative $\beta_6$ for down-hits implies that the sub-optimal trades exert a positive effect on $odds$. Based on the classification of shareholding market value, the Chinese stock markets investors are mainly divided into ordinary individual investors, professional individual investors and institutional investors according to the {\textit{Securities and Futures Investors Proper Management Measures of China Securities Regulatory Commission}} (2016). We can find out from Fig.~\ref{Fig:Suboptimal:tick:weight}, due to the fact that the sub-optimal orders are often particularly the orders which contain large trading volumes, and the ordinary individual investors have money constraints, that the submissions of the sub-optimal orders come from professional individual investors and institutional investors, so we can infer that professional individual investors and institutional investors play a positive role in the cooling effect when the stock price approaches its limits. The growth of professional individual investors and institutional investors is conducive to the long-term stability and healthy development of the Chinese stock markets.

\section{Identification of the cooling-off point}

\subsection{Common stocks}

We further add dummy variable $IR_{k-1}$ into the logit model as the explanatory variable to investigate where the magnet (cooling-off) effect emerges when the stock price approaches its limits. $IR_{k-1}^m$ for up-limit days is defined as follows,
\begin{equation}
 \label{Eq:logit:model:up:dummyvaribles:IR}
 IR_{k-1}^m = \left\{ \begin{array}{l}
 1,~~ {\mathrm{if}}\;R_{k-1} \geq m\%,~~~m=5,6,\cdots,9\\
 0,  ~~ {\mathrm{else}}\
 \end{array} \right.\
\end{equation}
where $R_{k-1} = \frac {p_{k-1}-P_i(T)}{P_i(T)}$, and $P_i(T)$ denotes the previous trading session's closing price of the limit-hitting day $i$.

Similarly, for down-limit days, we get,
\begin{equation}
 \label{Eq:logit:model:down:dummyvaribles:IR}
 IR_{k-1}^m = \left\{ \begin{array}{l}
 1, ~~~ {\mathrm{if}}\;R_{k-1} \leqslant -m\%,~~~-m=5,6,\cdots,9\\
 0,  ~~ {\mathrm{else}}\
 \end{array} \right.\
\end{equation}

Therefore, we rebuild the logit model as follows,
\begin{equation}
 \label{Eq:logit:regression:model:rebuild}
 \begin{split}
 X_k^{'}B = {\beta _0} + {\beta _1}\Delta {T_k}+ {\beta _2}{vol_{k-1}}*IBS_{k-1} + {\beta _3}{vol_{k-2}}*IBS_{k-2}+ {\beta _4}{vol_{k-3}}*IBS_{k-3}+ {\beta _5}{yield_{k-1}}\\
 + {\beta _6}{IR_{k-1}^m*yield_{k-1}}+{\beta _7}MKT_{k-1} + {\beta _8}MK{T_{k-2}} + {\beta _9}MK{T_{k-3}}+
 {\beta _{10}}spread_{k-1} + {\beta_{11}}depth_{k-1}
 \end{split}
\end{equation}

\setlength\tabcolsep{2pt}
\begin{table}[!ht]
  \centering
  \small
 \medskip
 \centering
\caption{\label{Tb:Statistics:logitmodel:estimates:conditional:probability} Summary statistics of parameter estimation from the logit regressions under the conditional probability. }
\begin{tabular}{ccccccccccccccccc}
\hline
\hline
\multicolumn{17}{l}{{\small{Panel A:  Parameter estimation of the logit regressions: Up-limit ($\alpha=0.05$)}}}\\
\hline
\multirow{2}{*}{Period} && \multirow{2}{*}{Model} && \multirow{2}{*}{$m$} && \multicolumn{3}{c}{$\beta_{5}$} && \multicolumn{3}{c}{$\beta_{6}$}&& \multicolumn{3}{c}{$\beta_{5}+\beta_{6}$ }\\
&& && &&\multicolumn{3}{c}{$\beta_{5}>0$, ~ $\beta_{5}<0$, ~ $\beta_{5}=0$}&& \multicolumn{3}{c}{$\beta_{6}>0$, ~ $\beta_{6}<0$, ~ $\beta_{6}=0$} &&\multicolumn{3}{c}{$ Mean(\beta_{5}+\beta_{6})$, ~ $ Odds(Mean)$  }\\
\cline{7-9}\cline{11-13}\cline{15-17}

Bullish && Up-limit  && 5 &&\multicolumn{3}{c}{167~~ 18494~~ 6409}&& \multicolumn{3}{c}{2654~~ 2321~~ 20095} && \multicolumn{3}{c}{-235.8 ~~ ~~   ~~~ ~~-21.0\%}\\
&& (N=25070)&& 6 &&  \multicolumn{3}{c}{161  ~  18519 ~ 6390}&& \multicolumn{3}{c}{3118 ~  2185  ~  19767} &&\multicolumn{3}{c}{-247.6 ~~ ~~  ~~~ ~~-21.9\%}\\
&& && 7 &&\multicolumn{3}{c}{159~~ 18574~~ 6337}&& \multicolumn{3}{c}{3632~~ 1908~~ 19530} &&\multicolumn{3}{c}{-254.5 ~~ ~~ ~~~ ~~-22.5\%}\\
&& &&8 &&\multicolumn{3}{c}{156~~ 18620~~ 6294}&& \multicolumn{3}{c}{3952~~ 1438~~ 19680} &&\multicolumn{3}{c}{ -257.6 ~~ ~~ ~~~ ~~-22.7\%}\\
&& &&9 &&\multicolumn{3}{c}{164~~ 18557~~ 6349}&& \multicolumn{3}{c}{3433~~~ 881~~~ 20756} &&\multicolumn{3}{c}{-255.8 ~~ ~~~ ~~~ ~~-22.6\%}\\\hline

Bearish && Up-limit &&5 &&\multicolumn{3}{c}{71~~ 16444~~ 6120}&& \multicolumn{3}{c}{2090~~ 2092~~ 18453} &&\multicolumn{3}{c}{-231.7 ~~ ~~  ~~~ ~~ -20.7\%}\\
&&(N=22635) &&6 &&\multicolumn{3}{c}{70~~ 16519~~ 6046}&& \multicolumn{3}{c}{2298~~ 2051~~ 18286} &&\multicolumn{3}{c}{-245.0 ~~ ~~ ~~~ ~~-21.7\%}\\
&& &&7 &&\multicolumn{3}{c}{69~~ 16606~~ 5960}&& \multicolumn{3}{c}{ 2653~~ 1816~~ 18166} &&\multicolumn{3}{c}{-255.2 ~~ ~~ ~~~ ~~-22.5\%}\\
&& &&8 &&\multicolumn{3}{c}{73~~ 16642~~ 5920}&& \multicolumn{3}{c}{3051~~ 1446~~ 18138} &&\multicolumn{3}{c}{-263.6 ~~ ~~ ~~~ ~~-23.2\%}\\
&& &&9 &&\multicolumn{3}{c}{75~~ 16615~~ 5945}&& \multicolumn{3}{c}{2788~~ 914~~ 18933} &&\multicolumn{3}{c}{-266.0 ~~ ~~~ ~~~ ~~-23.4\%}\\
\hline
\multicolumn{15}{l}{{\small{Panel B:  Parameter estimation of the logit regressions: Down-limit ($\alpha=0.05$)}}}\\
\hline
\multirow{2}{*}{Period} && \multirow{2}{*}{Model} && \multirow{2}{*}{$m$} && \multicolumn{3}{c}{$\beta_{5}$} && \multicolumn{3}{c}{$\beta_{6}$}&& \multicolumn{3}{c}{$\beta_{5}+\beta_{6}$ }\\
&& && &&\multicolumn{3}{c}{$\beta_{5}>0$, ~ $\beta_{5}<0$, ~ $\beta_{5}=0$}&& \multicolumn{3}{c}{$\beta_{6}>0$, ~ $\beta_{6}<0$, ~ $\beta_{6}=0$} &&\multicolumn{3}{c}{$ Mean(\beta_{5}+\beta_{6})$, ~ $ odds(Mean)$ }\\
\cline{7-9}\cline{11-13}\cline{15-17}
Bullish && Down-limit  && 5&&\multicolumn{3}{c}{11007~~ 38~~3708}&& \multicolumn{3}{c}{622~~ 2441~~ 11690} && \multicolumn{3}{c}{225.3 ~~ ~~   ~~~ ~~-20.2\%}\\
&& (N=14753)&& 6&&\multicolumn{3}{c}{11017~~ 41~~ 3695}&& \multicolumn{3}{c}{618~~ 2235~~ 11900} &&\multicolumn{3}{c}{225.9 ~~    ~~  ~~~ ~~-20.2\%}\\
&& && 7&&\multicolumn{3}{c}{11019~~ 39~~ 3695}&& \multicolumn{3}{c}{ 633~~ 1930~~ 12190} &&\multicolumn{3}{c}{226.1 ~~    ~~  ~~~ ~~-20.2\%}\\
&& &&8&&\multicolumn{3}{c}{11032~~ 38~~ 3683}&& \multicolumn{3}{c}{631~~ 1506~~ 12616} &&\multicolumn{3}{c}{229.9 ~~    ~~  ~~~ ~~-20.5\%}\\
&& &&9&&\multicolumn{3}{c}{11058~~ 41~~ 3654}&& \multicolumn{3}{c}{635~~ ~~ 936~~ ~~ 13182} &&\multicolumn{3}{c}{229.3 ~~ ~~ ~~~ ~~-20.5\%}\\\hline

Bearish && Down-limit  &&5 &&\multicolumn{3}{c}{17588~~ 14~~ 5665}&& \multicolumn{3}{c}{1394~~ 3425~~ 18488} && \multicolumn{3}{c}{183.6~~ ~~  ~~~ ~~-16.8\%}\\
&&(N=23267) &&6&&\multicolumn{3}{c}{17604~~ 11~~ 5652}&& \multicolumn{3}{c}{1280~~ 3333~~ 18654} &&\multicolumn{3}{c}{195.2~~    ~~  ~~~ ~~-17.7\%}\\
&& &&7&&\multicolumn{3}{c}{17594~~ 12~~ 5661}&& \multicolumn{3}{c}{ 1129~~ 3104~~ 19034} &&\multicolumn{3}{c}{202.8~~   ~~  ~~~ ~~-18.4\%}\\
&& &&8&&\multicolumn{3}{c}{17655~~ 13~~ 5599}&& \multicolumn{3}{c}{1222~~ 1609~~ 20436} &&\multicolumn{3}{c}{207.4  ~~  ~~ ~~~ ~~-18.7\%}\\
&& &&9&&\multicolumn{3}{c}{ 17630~~ 14~~ 5623}&& \multicolumn{3}{c}{995~~ ~~ 963~~ ~~ 21309} &&\multicolumn{3}{c}{204.5~~ ~~~  ~~~ ~~-18.5\%}\\
  \hline\hline
\end{tabular}
\end{table}

We rerun the logit regressions under conditional probability for 428,625 times and get the results in Table.~\ref{Tb:Statistics:logitmodel:estimates:conditional:probability}. The results include the numbers of significantly positive, negative at 5\% level. Panel A of Table~\ref{Tb:Statistics:logitmodel:estimates:conditional:probability} shows that when $m$ varies from 5 to 9(i.e., when the stock price approaches the limit-hitting price), the number of the significantly positive $\beta_5$ increases to some extent. $\beta_6$ is the parameter of $IR_{j-1}*yield_{k-1}$ and indicates that the price movement reaches a certain level. We find that the number of significantly negative (positive) $\beta_6$ decreases (increases) dramatically when $m$ varies from 5 to 9. Hence from the quantities transformation towards $\beta_5$, $\beta_6$, we conclude that there exists the cooling-off effect of up-limit models when the stock price approaches its limit-hitting price. Conversely, we find that $odds$ is decreasing as $m$ increases from 5 to 9, which indicates that there exists the cooling-off effect in another dimension. From the Panel B of Table~\ref{Tb:Statistics:logitmodel:estimates:conditional:probability}, we can also conclude that there exists the cooling-off effect for down-limit models.

\subsection{Specially treated stocks}

In contrast to stocks with 10\% price limit, we select ST and ST* shares with 5\% price limits on Chinese stock markets to investigate the magnet effect on different levels of price limits. Because trading activities of ST and *ST stocks in Chinese stock markets are not so active as normal listed stocks, we focus on pre-hit 20 ticks to investigate the performance of the market microstructure variables prior to its limits.

We rerun the logit regressions in Eq.~\ref{Eq:logit:regression:model:rebuild}. $IR_{k-1}^m$ is defined for the limit-up models as follows,
\begin{equation}
 \label{Eq:logit:model:5percent:up:dummyvaribles:IR}
 IR_{k-1}^m = \left\{ \begin{array}{l}
 1,~~{\mathrm{if}}\;R_{k-1} \geq m\%,~~~m =2.5,\cdots,4.5\\
 0,~~{\mathrm{else}}
 \end{array} \right.\
\end{equation}
where $R_{k-1} = \frac {p_{k-1}-P_i(T)}{P_i(T)}$.

Similarly, for the limit-down models, we get,
\begin{equation}
 \label{Eq:logit:model:5percent:down:dummyvaribles:IR}
 IR_{k-1}^m = \left\{ \begin{array}{l}
 1,~~{\mathrm{if}}\;R_{k-1} \leqslant -m\%,~~~m=2.5,\cdots,4.5\\
 0,~~{\mathrm{else}}
 \end{array} \right.\
\end{equation}

\setlength\tabcolsep{2pt}
\begin{table}[!ht]
  \centering
  \small
 \medskip
 \centering
\caption{\label{Tb:Statistics:logitmodel:ST:estimates}  Parameter estimations of the logit regressions about the ST, ST* stocks. }
\begin{tabular}{ccccccccccccccccc}
\hline
\hline
\multicolumn{17}{l}{{\small{Panel A: Parameter estimations of the logit regressions: Up-limit ($\alpha=0.05$)}}}\\
\hline
\multirow{2}{*}{Period} && \multirow{2}{*}{Model} && \multirow{2}{*}{$m$} && \multicolumn{3}{c}{$\beta_{5}$} && \multicolumn{3}{c}{$\beta_{6}$}&& \multicolumn{3}{c}{$\beta_{5}+\beta_{6}$ }\\
&& && &&\multicolumn{3}{c}{$\beta_{5}>0$, ~ $\beta_{5}<0$, ~ $\beta_{5}=0$}&& \multicolumn{3}{c}{$\beta_{6}>0$, ~ $\beta_{6}<0$, ~ $\beta_{6}=0$} &&\multicolumn{3}{c}{$ Mean(\beta_{5}+\beta_{6})$, ~ $ odds(Mean)$} \\
\cline{7-9}\cline{11-13}\cline{15-17}

Bullish && Up-limit  && 2.5 &&  \multicolumn{3}{c}{3~~ 1679~~ 1015}&& \multicolumn{3}{c}{386~~ 134~~ 2177} &&\multicolumn{3}{c}{-274.1 ~~  ~~  ~~~ ~~-24.0\%}\\
&&(N=2697) && 3 &&\multicolumn{3}{c}{4~~ 1676~~ 1017}&& \multicolumn{3}{c}{416~~ 141~~ 2140} &&\multicolumn{3}{c}{-286.3 ~~  ~~  ~~~ ~~-24.9\%}\\
&& &&3.5 &&\multicolumn{3}{c}{4~~ 1675~~ 1018}&& \multicolumn{3}{c}{405~~ 130~~ 2162} &&\multicolumn{3}{c}{-295.4 ~~  ~~  ~~~ ~~-25.6\%}\\
&& &&4 &&\multicolumn{3}{c}{4~~ 1672~~ 1021}&& \multicolumn{3}{c}{362~~ 112~~ 2223} &&\multicolumn{3}{c}{-292.8 ~~  ~~  ~~~ ~~-25.4\%}\\
&& &&4.5 &&\multicolumn{3}{c}{4~~ 1678~~ 1015}&& \multicolumn{3}{c}{248~~~ 81~~~ 2368} &&\multicolumn{3}{c}{-294.3 ~~  ~~  ~~~ ~~-25.5\%}\\\hline

Bearish && Up-limit &&2.5 &&\multicolumn{3}{c}{5~~ 1678~~ 1027}&& \multicolumn{3}{c}{265~~ 203~~ 2242} &&\multicolumn{3}{c}{-289.7 ~~  ~~  ~~~ ~~-25.2\%}\\
&& (N=2710)&&3 &&\multicolumn{3}{c}{5~~ 1686~~ 1019}&& \multicolumn{3}{c}{ 296~~ 198~~ 2216} &&\multicolumn{3}{c}{-277.4 ~~  ~~  ~~~ ~~-24.2\%}\\
&& &&3.5 &&\multicolumn{3}{c}{5~~ 1692~~ 1013}&& \multicolumn{3}{c}{310~~ 153~~ 2247} &&\multicolumn{3}{c}{-284.2 ~~  ~~  ~~~ ~~-24.7\%}\\
&& &&4 &&\multicolumn{3}{c}{6~~ 1687~~ 1017}&& \multicolumn{3}{c}{329~~ 127~~ 2254} &&\multicolumn{3}{c}{-283.7 ~~  ~~  ~~~ ~~-24.7\%}\\
&& &&4.5 &&\multicolumn{3}{c}{5~~ 1679~~ 1026}&& \multicolumn{3}{c}{255~~~ 89~~~ 2366} &&\multicolumn{3}{c}{-301.9 ~~  ~~  ~~~ ~~-26.1\%}\\
\hline
\multicolumn{15}{l}{{\small{Panel B: Parameter estimations of the logit regressions: Down-limit ($\alpha=0.05$)}}}\\
\hline
\multirow{2}{*}{Period} && \multirow{2}{*}{Model} && \multirow{2}{*}{$m$} && \multicolumn{3}{c}{$\beta_{5}$} && \multicolumn{3}{c}{$\beta_{6}$}&& \multicolumn{3}{c}{$\beta_{5}+\beta_{6}$ }\\
&& && &&\multicolumn{3}{c}{$\beta_{5}>0$, ~ $\beta_{5}<0$, ~ $\beta_{5}=0$}&& \multicolumn{3}{c}{$\beta_{6}>0$, ~ $\beta_{6}<0$, ~ $\beta_{6}=0$} &&\multicolumn{3}{c}{$ Mean(\beta_{5}+\beta_{6})$, ~ $ Odds(Mean)$}\\
\cline{7-9}\cline{11-13}\cline{15-17}
Bullish && Down-limit  && 2.5&&\multicolumn{3}{c}{1395~~ 1~~ 706}&& \multicolumn{3}{c}{164~~ 211~~ 1727} &&\multicolumn{3}{c}{238.5 ~~  ~~  ~~~ ~~-21.2\%}\\
&& (N=2102)&& 3&&\multicolumn{3}{c}{1402~~ 1~~ 699}&& \multicolumn{3}{c}{ 143~~ 228~~ 1731} &&\multicolumn{3}{c}{257.9 ~~  ~~  ~~~ ~~-22.2\%}\\
&& &&3.5&&\multicolumn{3}{c}{1396~~ 1~~ 705}&& \multicolumn{3}{c}{167~~ 199~~ 1736} &&\multicolumn{3}{c}{265.5 ~~  ~~  ~~~ ~~-23.3\%}\\
&& &&4&&\multicolumn{3}{c}{1400~~ 1~~ 701}&& \multicolumn{3}{c}{147~~ 156~~ 1799} &&\multicolumn{3}{c}{265.8 ~~  ~~  ~~~ ~~-23.3\%}\\
&& &&4.5&&\multicolumn{3}{c}{1406~~ 0~~ 696}&& \multicolumn{3}{c}{101~~ 103~~1898} &&\multicolumn{3}{c}{268.1 ~~  ~~  ~~~ ~~-23.5\%}\\\hline

Bearish && Down-limit  &&2.5&&\multicolumn{3}{c}{2063~~ 4~~ 1250}&& \multicolumn{3}{c}{187~~ 307~~ 2823} &&\multicolumn{3}{c}{225.3 ~~  ~~  ~~~ ~~-20.2\%}\\
&& (N=3317)&&3&&\multicolumn{3}{c}{2076~~ 3~~ 1238}&& \multicolumn{3}{c}{ 196~~ 321~~ 2800} &&\multicolumn{3}{c}{239.5 ~~  ~~  ~~~ ~~-21.3\%}\\
&& &&3.5&&\multicolumn{3}{c}{2062~~ 3~~ 1252}&& \multicolumn{3}{c}{212~~ 317~~ 2788} &&\multicolumn{3}{c}{265.9 ~~  ~~  ~~~ ~~-23.4\%}\\
&& &&4&&\multicolumn{3}{c}{ 2078~~ 3~~ 1236}&& \multicolumn{3}{c}{197~~ 266~~ 2854} &&\multicolumn{3}{c}{279.7 ~~  ~~  ~~~ ~~-24.4\%}\\
&& &&4&&\multicolumn{3}{c}{ 2077~~ 2~~ 1238}&& \multicolumn{3}{c}{161~~ 177~~ 2979} &&\multicolumn{3}{c}{276.2~~  ~~  ~~~ ~~-24.1\%}\\
  \hline\hline
\end{tabular}
\end{table}

We rerun the logit regressions for 64,956 times and present the results in Table~\ref{Tb:Statistics:logitmodel:ST:estimates}. We obtain the number of significantly positive, negative, and insignificantly estimates at 5\% level.
Table~\ref{Tb:Statistics:logitmodel:ST:estimates} reports the similar conclusion as Table~\ref{Tb:Statistics:logitmodel:estimates:conditional:probability}. We find that the logit regressions processed on the limit-hitting days with 5\% price limits with fitness testing and robustness testing are proved valid here. We further conclude that there exists the cooling-off effect when the stock price approaches its limits in Chinese stock markets for all different levels of price limits.

\section{Conclusion}

Due to the fact that the cooling-off effect or the magnet effect occurs at the intraday level, we  investigate the existence of the cooling-off effect (opposite to the magnet effect) by using high frequency data at the intraday level. For the lack of high frequency data on the Chinese stock markets, previous scholars studied the interday tendencies and interday properties to examine the effectiveness of the price limit trading rule. Through analyzing the intraday high-frequency data of all A-share common stocks traded on the Shanghai Stock Exchange and the Shenzhen Stock Exchange from 2000 to 2011, the findings of this article provide deeper insight into the mechanisms of the market structure and the effectiveness of the price limit trading rule.

In this paper, we conduct a study on the cooling-off effect (opposite to the magnet effect) from two aspects. Firstly we investigate the dynamics of five financial variables (trade size in lots, yield, volatility, bid-ask spread and limit-order book depth in lots) before limit hits under both bullish and bearish market states. And we find out the existence of the cooling-off effect before upper-limit hits and down-limit hits. Secondly, the logit model is adopted to investigate the cooling-effect before stock price hits its limits and the results demonstrate that cooling-off effect emerges for both up-limit hits and down-limit hits, and the cooling-off effect of down-limit hits is stronger than that of up-limit hits. The difference of the cooling-off effect between bullish period and bearish period is quite modest. This study provides a good explanation of the effectiveness of the price limit trading rule. Moreover, we examine the sub-optimal orders effect, and infer that professional individual investors and institutional investors play a positive role in the cooling-off effects. Generally speaking, the price limit trading rule adopted in the Chinese stock markets exerts a positive effect on restricting the investors' irrational behavior. A strengthened construction of professional individual investors and institutional investors will be conducive to the long-term stability and healthy development of the Chinese stock markets.

\section*{Acknowledgements}

This work was partially supported by the National Natural Science Foundation of China (71671066, 71501072), and the Fundamental Research Funds for the Central Universities (222201718006).

\bibliographystyle{elsarticle-num}
\bibliography{E:/Papers/Auxiliary/Bibliography}

\end{document}